\documentclass[twocolumn, sor, jor, amsmath,amssymb, reprint]{revtex4}

\usepackage{graphicx}% Include figure files
\usepackage{dcolumn}% Align table columns on decimal point
\usepackage{bm}% bold math
\usepackage{epstopdf}
\usepackage{color}
\usepackage{upgreek}
\begin{document}

%\preprint{}

	\title{Effect of edge disturbance on shear banding in polymeric solutions}
	\author{Seunghwan Shin}
	\affiliation{Department of Chemical Engineering and Materials Science, University of Minnesota, Minneapolis, Minnesota 55455, USA}
	\author{Kevin D. Dorfman}
	\email{dorfman@umn.edu}
	\affiliation{Department of Chemical Engineering and Materials Science, University of Minnesota, Minneapolis, Minnesota 55455, USA}
	\author{Xiang Cheng}
	\email{xcheng@umn.edu}
	\affiliation{Department of Chemical Engineering and Materials Science, University of Minnesota, Minneapolis, Minnesota 55455, USA}

\date{\today}

\begin{abstract}
Edge instabilities are believed to be one of the possible causes of shear banding in entangled polymeric fluids. Here, we investigate the effect of edge disturbance on the shear-induced dynamics of well-entangled DNA solutions. Using a custom high-aspect-ratio planar-Couette cell, we systematically measure the velocity profiles of sheared DNA samples at different distances away from the edge of the shear cell. Under a weak oscillatory shear with the corresponding Weissenberg number ($\mathrm{Wi}$) smaller than 1, where DNA solutions exhibit linear velocity profiles with strong wall slip, the penetration depth of the edge disturbance is on the order of the gap thickness of the shear cell, consistent with the behavior of Newtonian fluids. However, under a strong oscillatory shear with $\mathrm{Wi} > 1$ that produces shear-banding flows, the penetration depth is an order of magnitude larger than the gap thickness and becomes spatially anisotropic. Moreover, we find that the shear-banding flows persist deep inside the sheared sample, where the effect of edge disturbance diminishes. Hence, our experiments demonstrate an abnormally long penetration depth of edge disturbance and illustrate the bulk nature of shear-banding flows of entangled polymeric fluids under time-dependent oscillatory shear.   
\end{abstract}

%\pacs{}% insert suggested PACS numbers in braces on next line

\maketitle %\maketitle must follow title, authors, abstract and \pacs

\section{Introduction}
Under strong shear, an entangled polymeric fluid can develop heterogeneous flow profiles with multiple bands of different shear rates \cite{Wang11,Wang18}. Such shear-banding behavior has attracted great research interests in recent years. Although the experimental evidence for shear-banding in entangled polymeric fluids has accumulated in different polymer systems under various shear protocols, including steady and start-up shear and time-dependent oscillatory shear \cite{Tapadia06a,Ravindranath08,Boukany09a,Jaradat12,Shin17,Sato17,Goudoulas18}, the origin of these shear-banding flows is still under heated debate 
\cite{Hayes08,Hayes10,Hu12,Li13,Li14}. It is still controversial as to whether the observed shear banding flows arise from an underlying non-monotonic constitutive relation between shear stress and shear rate \cite{McLeish86,Doi88,Milner01}, confirmation of which would modify our current understanding of the nonlinear dynamics of entangled polymer chains \cite{Milner01}. To accommodate shear banding within the framework of the existing polymer theory, several alternative scenarios have been proposed, including strong flow-concentration coupling \cite{Fielding03,Cromer13}, localized
chain disentanglements \cite{Mohagheghi15,Mohagheghi16a} and long-lived transient instabilities
triggered by stress overshoot \cite{Cao12,Moorcroft13}. In particular, edge instabilities in the form of surface disturbances and edge fractures have been suggested as a possible cause of shear banding \cite{Schweizer08,Li13,Li14,Li15,Hemingway18}.

To mitigate the influence of edge instabilities, shear-banding experiments with sample edges wrapped in plastic films \cite{Boukany09a}, in large-aspect-ratio shear cells with small gaps \cite{Kirchenbuechler14,Boukany15,Sato17,Hemminger17} and using a special cone-partitioned-plate rheometer \cite{Ravindranath08b} have been conducted. However, it is still debatable whether these procedures truly eliminate the edge instabilities \cite{Li14}. Indeed, a recent numerical study showed that even a mild surface disturbance of shear-thinning polymeric fluids, which may go experimentally unnoticed, leads to strong secondary flows and apparent shear banding \cite{Hemingway18}. The secondary flows induced by the edge instability can penetrate deep into sheared samples up to 10--20 gap thicknesses $H$. These surprising results challenge not only the view of shear banding of polymeric fluids as a bulk phenomenon, but also the validity of conventional rheological characterization of strong shear-thinning polymeric fluids where the aspect ratio of shear cells $W/H$ is usually comparable to or smaller than the normalized penetration depth $L/H$. Inspired by this numerical study, we experimentally investigate the effect of edge instabilities on the shear banding profile of entangled polymeric fluids. %Therefore, it becomes critical to experimentally investigate the effect of edge instabilities on the shear banding profile of entangled polymeric fluids.   

Here, instead of minimizing edge instabilities, we use a custom-designed high-aspect-ratio planar shear cell to systematically probe the influence of edge disturbance on the shear-induced dynamics of entangled polymeric fluids. Specifically, we directly measure the shape variation of the velocity profile of entangled polymeric fluids as a function of the distance away from the edge. We find that the penetration depth of the edge disturbance is on the order of $H$ when the shear profile is linear, even with strong wall slip, similar to the behavior of Newtonian fluids. However, when shear-banding profiles develop, we observe a strong deviation from the bulk flow profile far away from the edge with a penetration depth an order of magnitude larger than $H$. The result is consistent with the prediction of Ref. \cite{Hemingway18}, even though a different boundary condition and shear protocol were adopted in simulations. Our results and these simulations suggest that a long penetration depth seems to be a generic feature for strong shear thinning polymer fluids, independent of boundary conditions and shear protocols. We furthermore find that the penetration depth is spatially inhomogeneous with a longer penetration along the flow direction. Under the condition of our experiments, the shear-banding flows persist deep inside entangled polymeric fluids when the edge effect vanishes, which thus eliminates edge disturbance as the origin of shear banding in our experiments. As such, our experiments reveal profound effects of boundary and edge on the velocity profiles of sheared complex fluids.

\begin{figure}
	\begin{center}
		\includegraphics[width=3.35in]{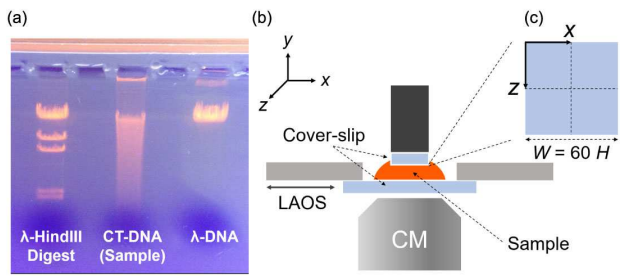}
		%select pdftexify command to run jpg or pdf files
	\end{center}
	\caption[Setup]{Sample and setup. (a) Gel electrphoresis of calf thymus DNA solution (center). The results from $\lambda$-DNA fragments (left, digested by HindIII restriction enzyme) and monodisperse $\lambda$-DNA (right, 48.5 kbp) are also added for comparison. All the DNA samples were prepared in the same TBE 2$\times$ buffer. (b) Schematic showing our custom planar-Couette cell (not to scale). CM: confocal microscope. A sheared sample is confined between two microscope coverslips with a fluid reservoir outside the cell. The gap thickness between the two coverslips is $H = 100$ $\upmu$m. (c) Top view of the top shear plate. The horizontal and vertical dashed lines indicate the two directions, along which we probe the edge effect.} \label{Figure1}
\end{figure}

\section{Experiments}
We used calf thymus DNA (double-stranded, average molecular weight 75 kbp, 4.9$\times 10^{7}$ g/mol, Affymetrix) as our model polymer, which is known to exhibit strong shear-banding flows under large amplitude oscillatory shear (LAOS) \cite{Boukany09a,Shin17,Goudoulas18}. Calf thymus DNA is less monodisperse compared with $\lambda$-DNA as shown in gel electrophoresis (Fig.~\ref{Figure1}a). The exact polydispersity depends on the specific method used for extracting DNA from calf thymus tissues and is not known from the vendor. Concentrated aqueous DNA solutions were prepared in $2\times$ TBE buffer (180 mM Tris base, 180 mM Boric acid, 5.6 mM EDTA). The buffer sufficiently screens the electrostatic interactions between DNA chains \cite{Hsieh08}. As a result, the DNA molecules behave similarly as neutral semi-flexible chains. We fixed the concentration of DNA at 8.3 mg/ml in this study, which is 160 times the overlap concentration. 

A standard rheological characterization of the DNA solution under small amplitude oscillatory shear was performed using a commercial cone-plate rheometer (AR-G2, TA Instruments) \cite{Shin17}. The plateau modulus of the solution is $G^{0}_{N} \approx 100$ Pa at 23 $^{\circ}$C. The average molecular weight of DNA molecules is calculated from the average chain length, $M = 7.5 \times 10^{4}$ bp $\times$ $650$ Da/bp $= 4.9 \times 10^{7}$ g/mol. Thus, the average number of entanglement points per chain, $Z$, can be estimated as $Z = (5/4)MG^{0}_{N}/(cRT) \approx 300$. Furthermore, the reciprocal of the overlap frequency at $G' = G''$ gives the reptation time $\tau_{d} = 900$ s. Although it is hard to estimate the effective Rouse time for highly polydisperse samples, we simply define the Rouse relaxation time of our DNA solutions as $\tau_{R} = \tau_{d}/3Z = 1$ s using the relation for monodisperse samples. The mesh size of the entangled network is about 90 nm. $R_{g}$ of our calf thymus DNA is 0.72 $\upmu$m, which is estimated based on $R_{g}$ of $\lambda$-DNA (48.5 kbp, 32 MDa) \cite{Teixeira07}. To track shear flows, we added a small amount of fluorescently-tagged polystyrene particles ($< 0.03$ wt$\%$) with radius 0.55 $\upmu$m in the solution for particle imaging velocimetry (PIV).  The tracer particle is more than 12 times larger than the entanglement length. The average distance between particles is 8.3 $\upmu$m, which is about $11R_{g}$.

Our setup is a custom planar-Couette cell, consisting of two parallel plates made of microscope coverslips (Fig.~\ref{Figure1}b) \cite{Shin17}. The coverslip glass is made of borosilicate from Thermo Fisher Scientific. The original dimension of the coverslips is $18 \times 18$ mm$^{2}$ and the thickness is labeled as $\#1$ (0.13 -- 0.17 mm). The coverslips were then cut to the right shape and size suitable as the top and bottom plates of the shear cell. The top plate is square with edge size $W = 6$ mm, whereas the bottom plate is circular with a much larger diameter of 12.8 mm. We washed the surface of the top and bottom plates with ethanol and water before each experiment to remove residual solutions and dust. Three differential screws located at the vertices of an equilateral triangle were used to adjust the level of the top plate relative to that of the bottom plate. The degree of parallelism was checked by measuring the distances between the top and bottom plates at the four corners of the square top plate using confocal microscopy. By finely tuning each screw with different amounts, one can achieve a good control of the level of the top plate with an accuracy of 1 $\upmu$m over 6 mm. The overall height of the gap can be lowered or increased by twisting the three screws together. More details about the design and the function of the setup can be further found in Ref. \cite{Lin14b}. 

During experiments, the top plate was held stationary, while the bottom plate was driven sinusoidally by a piezo actuator. We fixed the gap thickness between the top and the bottom plates at $H = 100$ $\upmu$m, equivalent to $\sim 140 R_{g}$, so that a high aspect ratio of $W/H = 60$ is maintained in our study. Note that due to the large mesh size and the long persistence length of DNA molecules, the cooperative diffusion length of DNA solutions is comparable to this gap size, which may lead to stronger concentration fluctuations between the gap than those in solutions of synthetic polymers. A DNA solution of volume $v = 15$ $\upmu$L was loaded into the shear cell before each experiment. Since $v$ is larger than the confined volume between the two shear plates, the solution outflows the edge of the top plate and forms a pinned contact line on the bottom plate. Hence, our experiments have a ``drown'' edge with a fluid reservoir outside the shear cell (Fig.~\ref{Figure1}b), a geometry frequently used in rheological measurements \cite{Hayes08,Hayes10,Vrentas91,Macosko94,Cohen06,Lin13,Lin14}. The shear cell was placed on a fast inverted confocal microscope for visualization of 3D flow profiles. 

%Three differential screws located at the vertices of an equilateral triangle are used to adjust the level of the top plate relative to that of the bottom plate. The degree of parallelism is checked by measuring the distances between the top and bottom plates at the four corners of the square top plate using confocal microscopy. By finely tuning each screw with different amounts, one can achieve a good control of the level of the top plate with an accuracy of 1 μm over 5 mm. The overall height of the gap can be lowered or increased by twisting the three screws together. Much more details about the design and the function of the device can be found in a previous publication by one of the authors (X.C.)

\begin{figure}
	\begin{center}
		\includegraphics[width=3.35in]{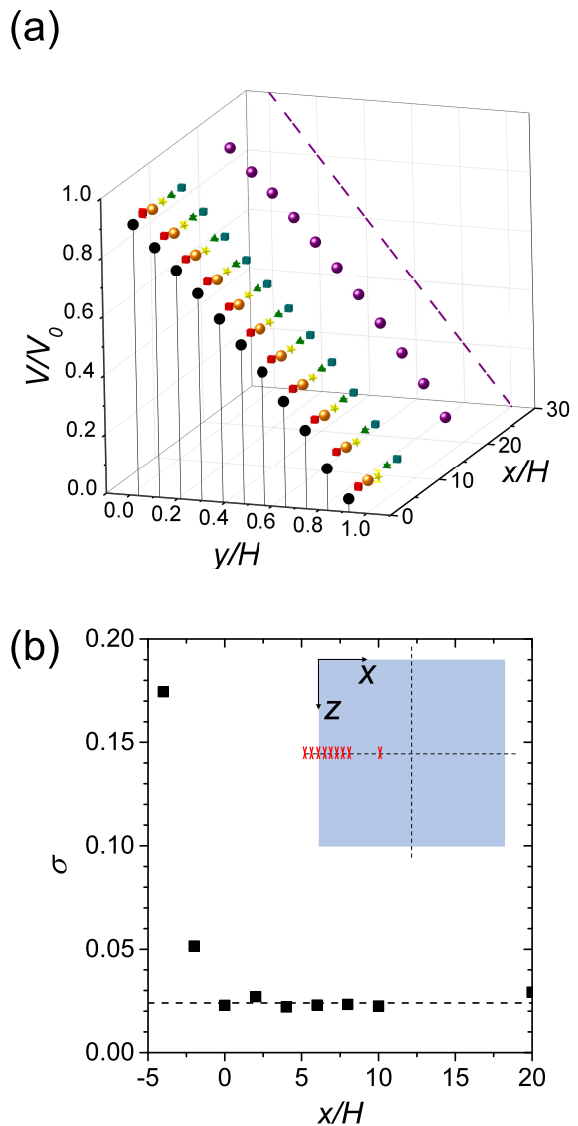}
		%select pdftexify command to run jpg or pdf files
	\end{center}
	\caption[Newtonian fluids]{Shear profiles of a glycerol/water mixture. Applied shear velocity amplitude $V_0 =3.77$ mm/s. Shear frequency $f=4.0$ Hz. $\mathrm{De} = 2 \pi \tau_{R} f = 25$. (a) Shear profiles, $V_x(x,y)$, at different locations $x$. $x$ and $y$ are normalized by $H$, whereas $V_x$ is normalized by $V_0$. From the front to back, $x/H= 0$, 2, 4, 6, 8, 10 and 20. The dashed line indicates the linear profile of a Newtonian fluid satisfying no-slip boundary conditions. (b) Standard deviation of the shape of shear profiles, $\sigma$, versus $x/H$. Intrinsic errors are indicated by the dashed line. Inset shows the top view of the top shear plate. Red crosses indicate the locations where the velocity profiles are measured.} \label{Figure2}
\end{figure}           

We define a coordinate system so that the flow, flow gradient and vorticity directions are along $x$, $y$ and $z$, respectively (Fig.~\ref{Figure1}c). The plane at $y=0$ indicates the position of the moving bottom plate, whereas the stationary top plate is at $y=H$. Since the bottom plate is driven sinusoidally with oscillatory velocity $V_0 \sin(2\pi f t)$, the flow at any location ($x$, $y$, $z$) within the shear cell simply follows $v(x, y, z, t) = V(x, y, z) \sin(2\pi f t + \delta)$. Here, $f$ is the shear frequency and $V_0$ is the applied velocity amplitude of the sinusoidal oscillatory shear, which is related to the amplitude of the displacement, $A_{0}$, via $V_{0} = 2\pi f A_{0}$. In our study, we fixed $A_0$ = 150 $\upmu$m and the corresponding shear strain $A_0/H = 1.5$, which gives $V_0 = 3.77$ mm/s at $f = 4$ Hz and 0.094 mm/s at $f = 0.1 Hz$. The phase shift $\delta$ is zero in our experiments. 

At fixed $x$ and $z$, $V(y)$ defines the velocity profile of sheared samples. We measure $V(y)$ at different locations along the bisector of the edge of the top plate either in the flow direction at $z=W/2$ (Fig.~\ref{Figure2}b inset) or in the vorticity direction at $x= W/2$ (Fig.~\ref{Figure5}b inset). We typically start the measurements from the edge and move gradually inward to the center of the sheared sample, although reversing the direction of experiments yields quantitatively the same results. For each $V(y)$ measurement, we take a video of four shearing cycles at a fixed $y$ and then scan different $y$ positions to obtain the entire velocity profile at given $x$ and $z$. These measurements are repeated three times, which are averaged to give the average velocity profile for the given sample. It takes $\sim$ 1 minute to obtain one average velocity profile at high $f$. Finally, three different samples are tested and averaged to yield the final results reported below. To remove the possible effect of sample loading, for each new sample, we preshear the sample at high shear rates for 15 min and let it rest for another 30 min before the start of the velocity profile measurements.   

Two dimensionless numbers can be constructed to quantify the dynamics of the DNA solution under oscillatory shear. The Weissenberg number ($\mathrm{Wi}$) of a shear is defined as $\mathrm{Wi} \equiv \tau_{R} V_{0}/H$. The Deborah number ($\mathrm{De}$) is defined as $De \equiv 2 \pi f \tau_{R}$, where $\tau_{R}$ is the Rouse relaxation time of DNA chains. For $\mathrm{Wi}$ and $\mathrm{De}$ defined based on the reptation time $\tau_d$, one can simply multiply the above definition of $\mathrm{Wi}$ and $\mathrm{De}$ by $3Z = 900$. 

\begin{figure}
	\begin{center}
		\includegraphics[width=3.35in]{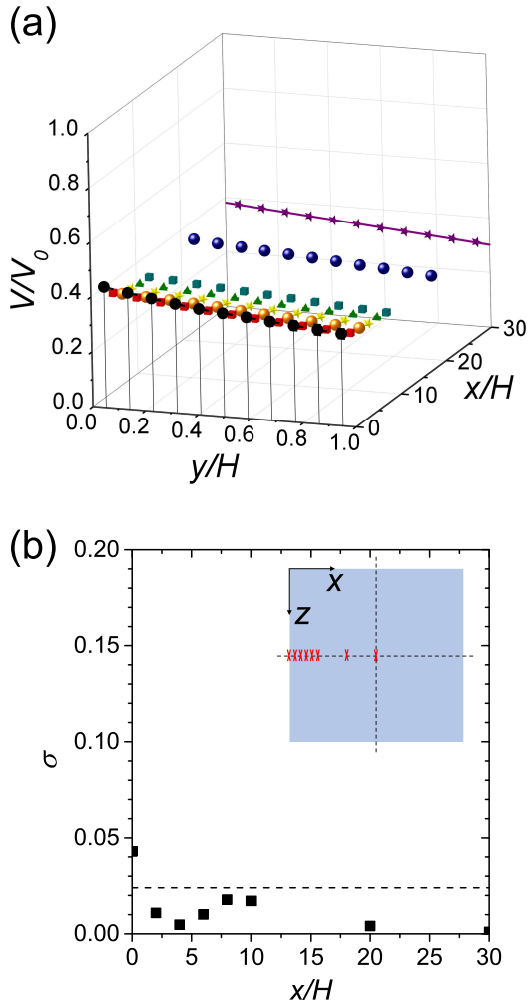}
		%select pdftexify command to run jpg or pdf files
	\end{center}
	\caption[Wall-slip of polymeric fluids]{Shear profiles of entangled DNA solutions at low $\mathrm{Wi}$. Applied shear velocity amplitude and frequency are $V_0 =0.094$ mm/s and $f=0.1$ Hz, respectively. $\mathrm{Wi} = 0.9$ and $\mathrm{De} = 0.6$. (a) Shear profiles, $V_x(x,y)$, at different locations $x$. From the front to back, $x/H= 0$, 2, 4, 6, 8, 10, 20 and 30. At all positions, the shear profiles are linear with significant wall slips. Linear fitting is applied to the profile at $x/H= 30$. (b) Standard deviation of the shape of shear profiles, $\sigma$, versus $x/H$. Intrinsic errors are indicated by the dashed line. Inset shows the top view of the top shear plate. Red crosses indicate the locations where the velocity profiles are measured.} \label{Figure3}
\end{figure} 
  
\section{Results and discussion}
To validate our experimental protocol, we first measure the edge effect on a glycerol/water mixture ($21/79$ wt$\%$), which is a Newtonian fluid with viscosity 1.7 mPa$\cdot$s. Velocity profiles of the mixture at different $x$ along the flow direction with $z=W/2$ are shown in Fig.~\ref{Figure2}a. As expected, $V(y)$ in the bulk of the sheared sample is linear. To quantify any deviation of the shape of the velocity profiles from the linear profile, we calculate the standard deviation of the shape variation of the velocity profiles as
\begin{equation}
\sigma(x,z) = \left[\frac{1}{H}\int_{0}^{H}\left(\frac{V(x,y,z)-\hat{V}(y)}{V_0}\right)^2 dy \right]^{1/2} 
\end{equation}
\begin{equation}
\hat{V}(y) = \lim_{W/H\to \infty} V\left(x \equiv \frac{W}{2}, y, z \equiv \frac{W}{2}\right), \nonumber
\end{equation} 
where $\hat{V}(y)$ is the velocity profile in the bulk without the influence of edge disturbance, which will be approximated in our study by a linear or piecewise linear fit to the velocity profile at $x = W/2$ and $z=W/2$ far away from the edge. For Newtonian fluids, $\hat{V}(y)$ is well known with $\hat{V}(y) = V_0(1 - y/H)$. We find that $\sigma(x,W/2)$ decreases sharply with $x$ and reaches a constant $c_0 = 0.024$ almost immediately when we move into the shear cell with $x \geq 0$ (Fig.~\ref{Figure2}b). Note that in our experiments, we also measure $V(x,y,z)$ outside the shear cell at negative $x$. The constant $c_0$ reflects intrinsic velocity fluctuations and errors of our PIV analysis, independent of edge disturbance. Accordingly, the penetration length, $L$, can be experimentally defined as the distance beyond which $\sigma$ plateaus and fluctuates around the noise threshold. In other words, $\sigma(x \ge L) \approx c_0$. For the glycerol/water mixture, $L < H$, consistent with the known result on the edge effect of Newtonian fluids with a fluid reservoir \cite{Vrentas91,Macosko94}.

Next, we measure the shape of the velocity profiles of the entangled DNA solutions. The velocity profiles of concentrated DNA solutions under LAOS have been well studied \cite{Wang11,Shin17}. Wang and co-workers suggest that the flow behaviors of entangled polymeric fluids can be predicted based on $2b_{max}/H$ and $\mathrm{Wi}$, where $b_{max}$ is the maximal slip length \cite{Wang11}. For a $1\%$ water-based DNA solution of an average chain length comparable to our system, Boukany {\it et al.} shows $2b_{max}/H = 136$, where $H = 1$ mm in their study \cite{Boukany08}. Thus, we estimate $2b_{max}/H \approx 1300$ in our study with $H = 0.1$ mm. The transition Weissenberg number between wall slip and shear banding is given by $\mathrm{Wi}_{ws-sb}^{Rp} = 1 + 2 b_{max}/H \approx 1300$, where $\mathrm{Wi}^{Rp}$ is estimated based on the reptation dynamics \cite{Wang11}. The transition Weissenberg number based on Rouse dynamics can be simply calculated as $\mathrm{Wi}_{ws-sb} = \mathrm{Wi}_{ws-sb}^{Rp} /3Z = 1.4$. Below $\mathrm{Wi}_{ws-sb}$, one expects to observe wall-slip dominated shear profiles, whereas above $\mathrm{Wi}_{ws-sb}$, shear banding occurs.

\begin{figure}
	\begin{center}
		\includegraphics[width=3.35in]{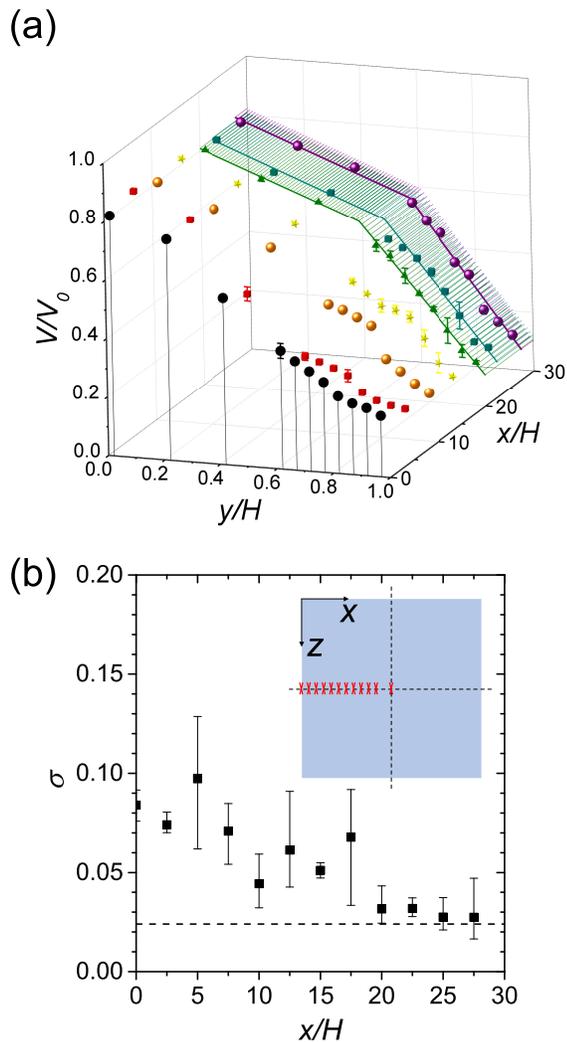}
		%select pdftexify command to run jpg or pdf files
	\end{center}
	\caption[Shear banding of polymeric fluids]{Shear profiles of entangled DNA solutions at high $\mathrm{Wi}$. Applied shear velocity amplitude and frequency are $V_0 =3.77$ mm/s and $f=4.0$ Hz, respectively. $\mathrm{Wi}=38$ and $\mathrm{De} = 25$. (a) From the front to back, $x/H= 0$, 10, 15, 20, 22.5 and 27.5. Piecewise linear fittings are applied to the shear-banding profiles deep inside the sheared sample. (b) Standard deviation of the shape of shear profiles, $\sigma$, versus $x$. Intrinsic errors are indicated by the dashed line. Inset shows the top view of the top shear plate. Red crosses indicate the locations where the velocity profiles are measured.} \label{Figure4}
\end{figure}  

In our experiments, at low $\mathrm{Wi} = 0.9 < \mathrm{Wi}_{ws-sb}$, we indeed observe the linear profile with strong wall slips in the bulk of the shear cell (Fig.~\ref{Figure3}a). Near the edge, a deviation from the bulk linear velocity profile can be found. Quantitatively, the standard deviation of the shape variation along the flow direction, $\sigma(x,W/2)$, decreases near the edge and plateaus when $x \gtrsim H$ (Fig.~\ref{Figure3}b). Notice that when calculating $\sigma$, we fit the velocity profile of the sheared entangled DNA solution at the center of the shear cell $x=W/2$ and $z=W/2$ using $\hat{V}(y) = V_0(H+l_t-y)/(H+l_t+l_b)$, where the slip lengths at the top and bottom plates, $l_t$ and $l_b$, are two fitting parameters. Hence, our experiments show that although the entangled DNA solution shows strong viscoelasticity and shear thinning \cite{Shin17}, the penetration depth $L$ is still on the order of $H$ when the velocity profile is linear (albeit with strong wall slips), quantitatively similar to the edge effect on Newtonian fluids.      

\begin{figure}
	\begin{center}
		\includegraphics[width=3.35in]{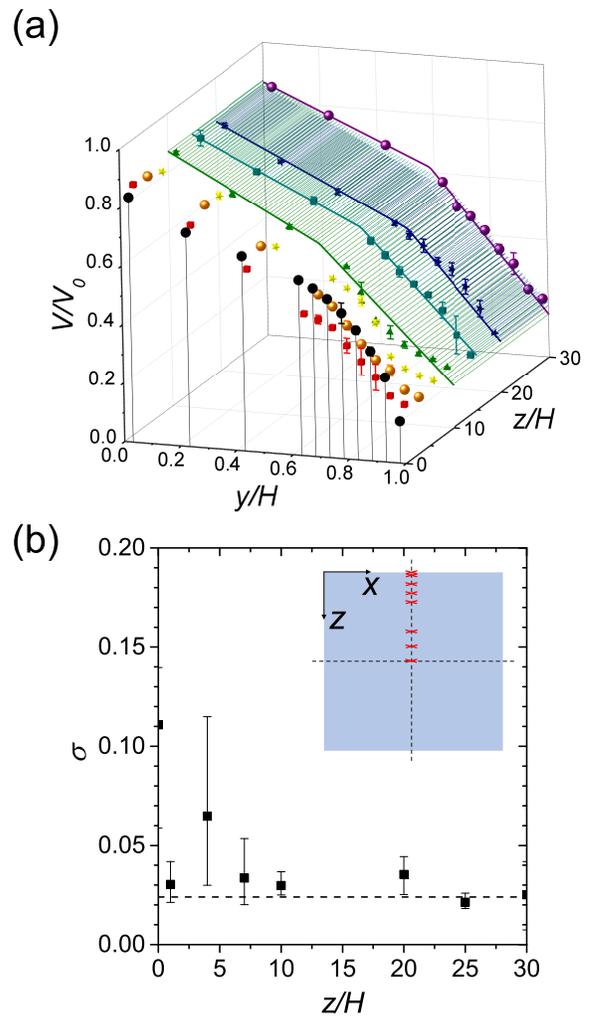}
		%select pdftexify command to run jpg or pdf files
	\end{center}
	\caption[Shear banding of polymeric fluids in voriticity]{Shear profiles of entangled DNA solutions along the vorticity direction at high $\mathrm{Wi}$. Applied shear velocity amplitude and frequency are same as those in Fig.~\ref{Figure4}, $V_0 =3.77$ mm/s and $f=4.0$ Hz. $\mathrm{Wi} = 38$ and $\mathrm{De} = 25$. (a) From the front to back, $z/H= 0$, 1, 4, 7, 10, 15, 20, 30. Piecewise linear fittings are applied to the shear-banding profiles deep inside the sheared sample. (b) $\sigma (z)$ obtained by comparing each profile to the piecewise linear fitting of the shear-banding profile at $z/H=30$. Inset shows the top view of the top shear plate. Red crosses indicate the locations where the velocity profiles are measured.} \label{Figure5}
\end{figure}  

At high $\mathrm{Wi} = 38>\mathrm{Wi}_{ws-sb}$, we also verify the existence of shear-banding flows deep in the sheared entangled DNA solution. The velocity profiles change substantially with $x$ near the edge due to edge disturbance (Fig.~\ref{Figure4}a). In contrast to the case of low $\mathrm{Wi}$, the decrease of $\sigma(x,W/2)$ along the flow direction is much slower with increasing $x$. The shape of the velocity profiles gradually stabilizes over a surprisingly long distance of $\sim 20H$ (Fig.~\ref{Figure4}b). Here, to calculate $\sigma$, we obtain $\hat{V}(y)$ by fitting the shear profile at $x=W/2$ and $z=W/2$ using piecewise linear lines. Thus, our entangled DNA solution displays a penetration depth one order magnitude larger than $H$, qualitatively agreeing with the numerical finding \cite{Hemingway18}. However, the long penetration was observed along the vorticity direction, instead of along the flow direction, in simulations. Moreover, different from simulations, the shear-banding profile persists in the bulk of the sheared sample in our experiments. Such differences may arise from different boundary conditions and shear protocols used in simulations and experiments. Indeed, instead of reducing the degree of shear banding, the velocity profile becomes more heterogeneous deeper inside the sample in our experiments (Fig.~\ref{Figure4}a). This observation eliminates the edge disturbance as the possible origin of shear-banding flows in our LAOS experiments.

\begin{figure}
	\begin{center}
		\includegraphics[width=3.35in]{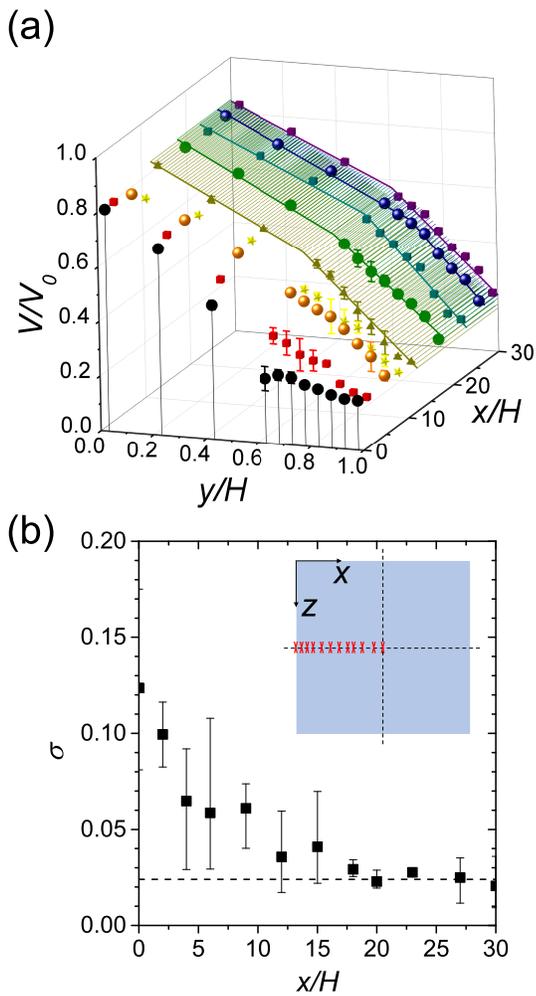}
		%select pdftexify command to run jpg or pdf files
	\end{center}
	\caption[Shear banding of polymeric fluids with larger volume, $v=40 mu$L]{Shear profiles of entangled DNA solutions with larger fluid reservoir at high $\mathrm{Wi} = 38$ and $\mathrm{De} = 25$.  Applied shear velocity amplitude and frequency are the same as those in Fig.~\ref{Figure4}. $\mathrm{Wi}=38$. Sample volume $v=40$ $\upmu$L. (a) From the front to back, $x/H= 0$, 2, 6, 9, 12, 18, 23, 27, and 30. Piecewise linear fittings are applied to the shear-banding profiles deep inside the sheared sample. (b) Standard deviation of the shape of shear profiles, $\sigma$, versus $x$. Intrinsic errors are indicated by the dashed line. Inset shows the top view of the top shear plate. Red crosses indicate the locations where the velocity profiles are measured.} \label{Figure6}
\end{figure}  

We also investigate the influence of edge disturbance along the vorticity direction. Specifically, we measure the velocity profiles at different locations $z$ along the bisector of the edge of the top plate normal to the vorticity direction at $x = W/2$ (Fig.~\ref{Figure5}b inset). Figure \ref{Figure5}a shows the velocity profiles at different $z$ for the entangled polymer solution at high $\mathrm{Wi}$, where strong shear-banding flows are observed deep inside the bulk. Shear banding is again less obvious near the edge of the shear cell, similar to that along the flow direction. However, although the penetration depth along the flow direction is $\sim 20H$ (Fig.~\ref{Figure4}b), the penetration depth along the vorticity direction appears to be much smaller on the order of $H$. The finding contradicts the numerical result of Hemingway and Fielding, where a long penetration depth is found along the vorticity direction \cite{Hemingway18}. Since the simulation assumes a translational invariance along the flow direction that eliminates the existence of the $x$ edges, it is not clear if spatially inhomogeneous penetration depths can be detected in the numerical model adopted in \cite{Hemingway18}. Our result has some interesting implications for conventional rheology measurements. Since the normal direction of the air-fluid interface of sheared samples in a cone-plate rheometer is also along the vorticity direction, edge disturbance may not strongly influence the bulk velocity profiles. However, it should be emphasized that the curvature of the interface in rotational rheometers can also trigger edge instabilities \cite{Macosko94,Li14}, a factor that cannot play a role in our planar shear cell.         
 
Lastly, we also study the effect of the size of fluid reservoirs on the change of the velocity profiles. A large volume of the DNA solution of $v=40$ $\upmu$L is used in this experiment, which gives rise to a significantly larger fluid reservoir compared to those experiments with $v=15$ $\upmu$L solutions. Figure \ref{Figure6} shows the velocity profiles along the flow direction at high $\mathrm{Wi}$. The shear condition is the same as that used in Fig.~\ref{Figure4}. The results are qualitatively similar as those shown in Fig.~\ref{Figure4} too; $\sigma$ decreases near the edge and plateaus around $x = 10 - 15H$, which again suggests an abnormal long penetration depth of $L \sim 10 - 15H$. Quantitatively, it seems that a larger fluid reservoir leads to a smaller penetration depth, consistent with the expectation for Newtonian fluids \cite{Vrentas91,Macosko94}.  

Although inspired by the work of Hemingway and Fielding, our experiments are different from the simulations in two key aspects, which affect the direct comparison between experimental and numerical results. First, we apply large amplitude oscillatory shear (LAOS) instead of steady shear in our experiments. Shear-banding in time-dependent flows may have different origins from steady shear-banding \cite{Fielding16}. Hence, although our experiments exclude edge instabilities as the origin of shear-banding in LAOS flows, one should be cautious when extending the same conclusion to steady shear-banding flows \cite{Ravindranath08}. Second, the lateral boundary of our shear cell is different from that of the simulations, where the sheared samples are completely confined between two parallel shear plates. This difference likely explains why we do not observe strong shear banding flows near the edge induced by the edge disturbance. Bulk shear-banding flows emerge only when the effect of edge disturbance diminishes. The shear-banding profiles of entangled DNA solutions are well-established at $x \approx 20H$ away from the edge of the cell.

\section{Conclusions}

In conclusion, we systematically investigated the effect of edge disturbance on the velocity profiles of highly entangled DNA solutions. In particular, we measured the penetration depth of edge disturbance. Under weak shear with linear shear profiles, the solutions exhibit a short penetration depth comparable to the gap thickness of the shear cell, consistent with our understanding based on Newtonian fluids. However, under strong shear with shear-banding flows, the penetration depth is one order of magnitude larger than the gap thickness along the flow direction, confirming the existence of an abnormally long penetration of edge disturbance \cite{Hemingway18}. In addition, we found that the penetration depth is anisotropic. The influence of edge disturbance is significantly deeper along the flow direction than along the vorticity direction. Moreover, a larger fluid reservoir results in a slightly shorter penetration depth, a feature that may be exploited in standard rheological tests of entangled polymeric fluids. Finally, we verified that LAOS exerted in our experiments gives rise to true bulk shear-banding flows without the influence of edge disturbance. Our work illustrates the profound effects of edge disturbance on the sheared dynamics of entangled polymer fluids.

\begin{acknowledgments}

We thanks Suzanne Fielding for comments on an early version of this manuscript. The work was supported by NSF CBET-1700771. 

\end{acknowledgments}

%\bibliography{}

\end{document}